\documentclass[prl,twocolumn,showpacs]{revtex4-1}
\usepackage{graphicx}

\newcommand{\andauthor}{and }
\newcommand{\etal}{\textit{et al.} }

\newcommand{\putauthor}[2]{#1 #2}

\newcommand{\volstyle}[1]{\textbf{#1}}
\newcommand{\joustyle}[1]{#1}
\newcommand{\yearstyle}[1]{(#1)}
\newcommand{\pagstyle}[1]{#1}

\newcommand{\journal}[4]{\joustyle{#1} \volstyle{#2}, \pagstyle{#4} \yearstyle{#3}.}
\newcommand{\tapj}[3]{\journal{Ap.\ J.}{#1}{#2}{#3}}
\newcommand{\tapjl}[3]{\journal{Ap.\ J. Lett.}{#1}{#2}{#3}}
\newcommand{\tapjs}[3]{\journal{Ap.\ J. Suppl.}{#1}{#2}{#3}}
\newcommand{\taanda}[3]{\journal{Astron.\ Astrophys.}{#1}{#2}{#3}}
\newcommand{\tprc}[3]{\journal{Phys.\ Rev.\ C}{#1}{#2}{#3}}
\newcommand{\tnpa}[3]{\journal{Nucl.\ Phys.}{A#1}{#2}{#3}}
\newcommand{\tadndt}[3]{\journal{At.\ Data Nucl.\ Data Tables}{#1}{#2}{#3}}
\newcommand{\trmp}[3]{\journal{Rev.\ Mod.\ Phys.}{#1}{#2}{#3}}
\newcommand{\tphysrep}[3]{\journal{Phys.\ Rep.}{#1}{#2}{#3}}

\begin{document}


\title{Solution of the $\alpha$-potential mystery in the $\gamma$-process and its impact on the Nd/Sm ratio in meteorites}


\author{Thomas Rauscher}
\email{t.rauscher@herts.ac.uk}

\affiliation{Centre for Astrophysics Research, School of Physics, Astronomy and Mathematics, University of Hertfordshire, College Lane, Hatfield AL10 9AB, United Kingdom}
\affiliation{Institute for Nuclear Research, Hungarian Academy of Science, 4001 Debrecen, Hungary}
\affiliation{Department of Physics, University of Basel, Klingelbergstrasse 82, CH-4056 Basel, Switzerland}


\date{\today}

\begin{abstract}
The $^{146}$Sm/$^{144}$Sm ratio in the early solar system has been constrained by Nd/Sm isotope ratios in meteoritic material. Predictions of $^{146}$Sm and $^{144}$Sm production in the $\gamma$-process in massive stars are at odds with these constraints and this is partly due to deficiences in the prediction of the reaction rates involved. The production ratio depends almost exclusively on the ($\gamma$,n)/($\gamma$,$\alpha$) branching at $^{148}$Gd. A measurement of $^{144}$Sm($\alpha$,$\gamma$)$^{148}$Gd at low energy had discovered considerable discrepancies between cross section predictions and the data. Although this reaction cross section mainly depends on the optical $\alpha$+nucleus potential, no global optical potential has yet been found which can consistently describe the results of this and similar $\alpha$-induced reactions at the low energies encountered in astrophysical environments. The untypically large deviation in $^{144}$Sm($\alpha$,$\gamma$) and the unusual energy dependence can be explained, however, by low-energy Coulomb excitation which is competing with compound nucleus formation at very low energies. Considering this additional reaction channel, the cross sections can be described with the usual optical potential variations, compatible with findings for (n,$\alpha$) reactions in this mass range. Low-energy ($\alpha$,$\gamma$) and ($\alpha$,n) data on other nuclei can also be consistently explained in this way. Since Coulomb excitation does not affect $\alpha$-emission, the $^{148}$Gd($\gamma$,$\alpha$) rate is much higher than previously assumed. This leads to small $^{146}$Sm/$^{144}$Sm stellar production ratios, in even more pronounced conflict with the meteorite data.
\end{abstract}


\maketitle


The astrophysical $\gamma$-process synthesizes proton-rich nuclides through sequences of photodisintegrations of pre-existing seed material. It occurs in explosive Ne/O burning in core-collapse supernova (ccSN) explosions of massive stars \cite{woohow78,rhhw02}. This site was supposed to be the main source of the p-nuclides, i.e., naturally occurring, proton-rich nuclei which cannot be produced in the s- and r-process \cite{raupreview,arngorp}.
A recent investigation has shown that also type Ia supernovae (SNIa) may be a viable site for the $\gamma$-process  \cite{travWD,travWDproc}, although previous simulations had not been successful \cite{howmey,nom}.

The $\gamma$-process produces both $^{146}$Sm and $^{144}$Sm, the production ratio $\mathcal{R}\equiv P_{146}/P_{144} \propto \lambda_{\gamma \mathrm{n}}/\lambda_{\gamma \alpha}=R_{\gamma \mathrm{n}}/R_{\gamma \alpha}$ depends on the stellar ($\gamma$,n) and ($\gamma$,$\alpha$) rates of $^{148}$Gd, denoted by $\lambda_{\gamma \mathrm{n}}$ and $\lambda_{\gamma \alpha}$, respectively, or alternatively on the ratios of the reactivities, denoted by $R_{\gamma \mathrm{n}}$ and $R_{\gamma \alpha}$ \cite{woohow}. This ratio is of particular interest because it was suggested that surviving $^{146}$Sm may be detected in the solar system and used for cosmochronometry \cite{aud}. No live $^{146}$Sm has been found to date but at least the signature of its in-situ decay in meteorites is believed to be seen, from which the isotope ratio at the closure of the solar system can be inferred \cite{prinz,harper,kino,raupreview}.

There are still large uncertainties involved in determining the production ratio, both from the side of astrophysical models and from nuclear physics. To better constrain the nuclear uncertainties $^{144}$Sm($\alpha$,$\gamma$)$^{148}$Gd was measured in a pioneering, difficult experiment \cite{sm144}. Since the stellar $\alpha$-capture reactivity $R_{\alpha \gamma}$ is dominated by the ground state (g.s.) transition \cite{woohow,sensipaper}, the laboratory value can be converted to the stellar ($\gamma$,$\alpha$) reactivity $R_{\gamma \alpha}$ by applying detailed balance \cite{woohow,raureview}. Although the astrophysically relevant energy range of 9 MeV and below \cite{energywindows} could not be reached, the lowest datapoint at 10.2 MeV already showed a strong deviation from predictions. Using an optical $\alpha$+nucleus potential with an energy-dependent part fitted to reproduce the data, a stellar ($\gamma$,$\alpha$) rate was derived which was lower by an order of magnitude than previous estimates \cite{sm144} (see Table \ref{table}). This led to a strongly increased $\mathcal{R}$.

This result shed doubts on the prediction of ($\gamma$,$\alpha$) rates at $\gamma$-process temperatures and triggered a number of experimental and theoretical studies. Due to the tiny reaction cross sections, however, data is still scarce in the relevant mass region (at neutron numbers $N\geq 82$) and close to astrophysical energies. A comparison of predictions to data at higher energy often is irrelevant because the cross sections depend not only on the $\alpha$-widths, as they do at low energy \cite{sensipaper}. To calculate the reaction cross sections in the Hauser-Feshbach model \cite{haufesh}, so-called optical potentials -- describing the effective interaction between projectile and target nucleus -- have to be used in the numerical solution of the Schr\"odinger equation. Many local and global optical $\alpha$+nucleus potentials have been derived, using elastic scattering at higher energy, reaction cross sections, and theoretical considerations (like folding potentials), e.g., see \cite{kis09,mohradndt} and references therein. None of the potentials are able to describe the existing ($\alpha$,$\gamma$) and ($\alpha$,n) data consistently, yet.
Rather, a seemingly confusing picture arises. Some of the low-energy data are described well, the majority of cases find deviations increasing with decreasing energies but never exceeding overprediction factors $2-3$, and then there is the $^{144}$Sm($\alpha$,$\gamma$) case with its large deviation of more than an order of magnitude. Also the energy dependence of the $^{144}$Sm($\alpha$,$\gamma$) data is peculiar and cannot be reproduced by any prediction (unless fitted to the data). The only common factor seems to be that the predictions using the standard optical potential \cite{mcf} are either close to the data or considerably higher.

Instead of attempting to solve the discrepancy by modifying the nuclear interaction potentials alone, another approach is suggested here.
The low-energy deviations and their variation from one nucleus to another can be explained by the action of an additional reaction channel which was not considered in the calculations, such as a direct inelastic channel (direct elastic scattering is included in the usual optical potentials \cite{sat83}). In the picture of the optical model, this channel would divert part of the impinging $\alpha$-flux away from the compound nucleus formation channel and thus lead to fewer compound nuclei at a given projectile flux. In the experiment this is seen as smaller reaction yield. Coulomb excitation (Coulex) is such a reaction mechanism. It has been used extensively to study nuclear structure and it is well known that Coulex cross section can be comparable to or larger than compound reaction cross sections at several tens of MeV. At lower energies they are commonly assumed to be negligible compared to the compound formation cross section. It can be shown, however, that for intermediate and heavy nuclei the Coulex cross section $\sigma^\mathrm{Coulex}$ declines more slowly with decreasing energy than the compound formation cross section $\sigma^\mathrm{form}$, due to the Coulomb barrier. Consequently, the Coulex cross section can become comparable to or even exceed the compound formation cross section close to the astrophysical energy range and below, even when it has been negligible at intermediate energy.

\begin{figure}[h]
\begin{center}
\includegraphics[angle=-90,width=0.8\columnwidth]{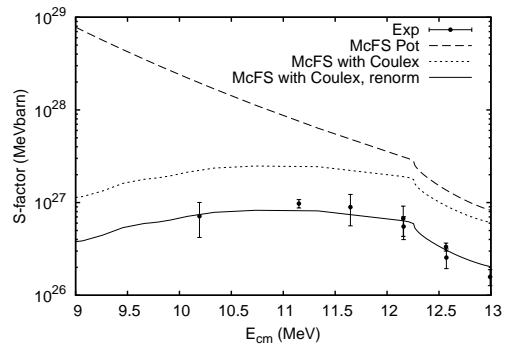}
\end{center}
\caption{Experimental $S$ factors for $^{144}$Sm($\alpha$,$\gamma$)$^{148}$Gd \cite{sm144} (Exp) are compared to predictions using the standard potential \cite{mcf} (McFS, dashed line), using the same potential but corrected for Coulomb excitation (dotted line), and the Coulex corrected prediction with the $\alpha$-width divided by a constant factor of 3 (full line). The astrophysically relevant energy is about $8-9$ MeV \cite{energywindows}.}
\label{fig:sm144}
\end{figure}

A straightforward way to include the diversion of $\alpha$-flux from the compound formation channel in the cross section calculation is to use a modified compound formation cross section (this can simply be implemented by using modified $\alpha$-transmission coefficients in the \emph{entrance} channel) $\sigma_\ell^\mathrm{form,mod}=f_\ell \sigma^\mathrm{form}$, with
\begin{equation}
\label{eq:coulextrans}
f_\ell=\frac{\sigma_\ell^\mathrm{form}}{\sigma_\ell^\mathrm{form}+\sigma_\ell ^\mathrm{Coulex}}
\end{equation}
for each partial wave $\ell$.
The Coulex cross section can be calculated, e.g., by \cite{alder}
\begin{eqnarray}
&&\sigma_\ell^\mathrm{Coulex} \propto \nonumber \\
&&B(\mathrm{E} \mathcal{L}) \sum_{\ell_\mathrm{f}} {\left\{ \left(2 \ell_\mathrm{f} +1 \right) \left| \int\limits_0^\infty {F_{\ell_\mathrm{f}}(k_\mathrm{f}r) r^{-\mathcal{L}-1} F_\ell (kr) \, dr} \right| \right\} },
\end{eqnarray}
using regular Coulomb wave functions $F_\ell (kr)$, $F_{\ell_\mathrm{f}}(k_\mathrm{f}r)$ at initial and final $\alpha$-energies, respectively. The transition strengths for electric multipole emission of multipolarity $\mathcal{L}$ are given by $B(\mathrm{E} \mathcal{L})$. The results shown here are for the dominant multipolarity $\mathcal{L}=2$, i.e., E2 transitions and were obtained using a newly developed Hauser-Feshbach code, called SMARAGD \cite{smaragd}.

In astrophysical investigations, often the S-factor $S(E)=\sigma E \exp(2\pi \eta)$ is given rather than the reaction cross section $\sigma$, with the exponential including the Sommerfeld parameter $\eta$ accounting for the Coulomb barrier penetration.
Figure \ref{fig:sm144} shows how the S-factor of $^{144}$Sm($\alpha$,$\gamma$)$^{148}$Gd is changed by inclusion of Coulex while still using the standard potential \cite{mcf}. The energy dependence of the data is now accurately reproduced but the absolute value is still too high. It was assumed in the calculation, however, that the optical potential used accurately describes compound formation in the absence of Coulex. This does not have to be the case, though, there may still be an additional energy dependence which has to be determined independently. The data can be described well by renormalizing the $\alpha$-widths obtained with the standard potential, as also shown in Fig.\ \ref{fig:sm144}. The required factor of 1/3 is well in line with the typical deviations found for other reactions involving $\alpha$ particles at low energy, for which no Coulex occurs (e.g., in (n,$\alpha$) reactions).

\begin{figure}[h]
\begin{center}
\includegraphics[angle=-90,width=0.65\columnwidth]{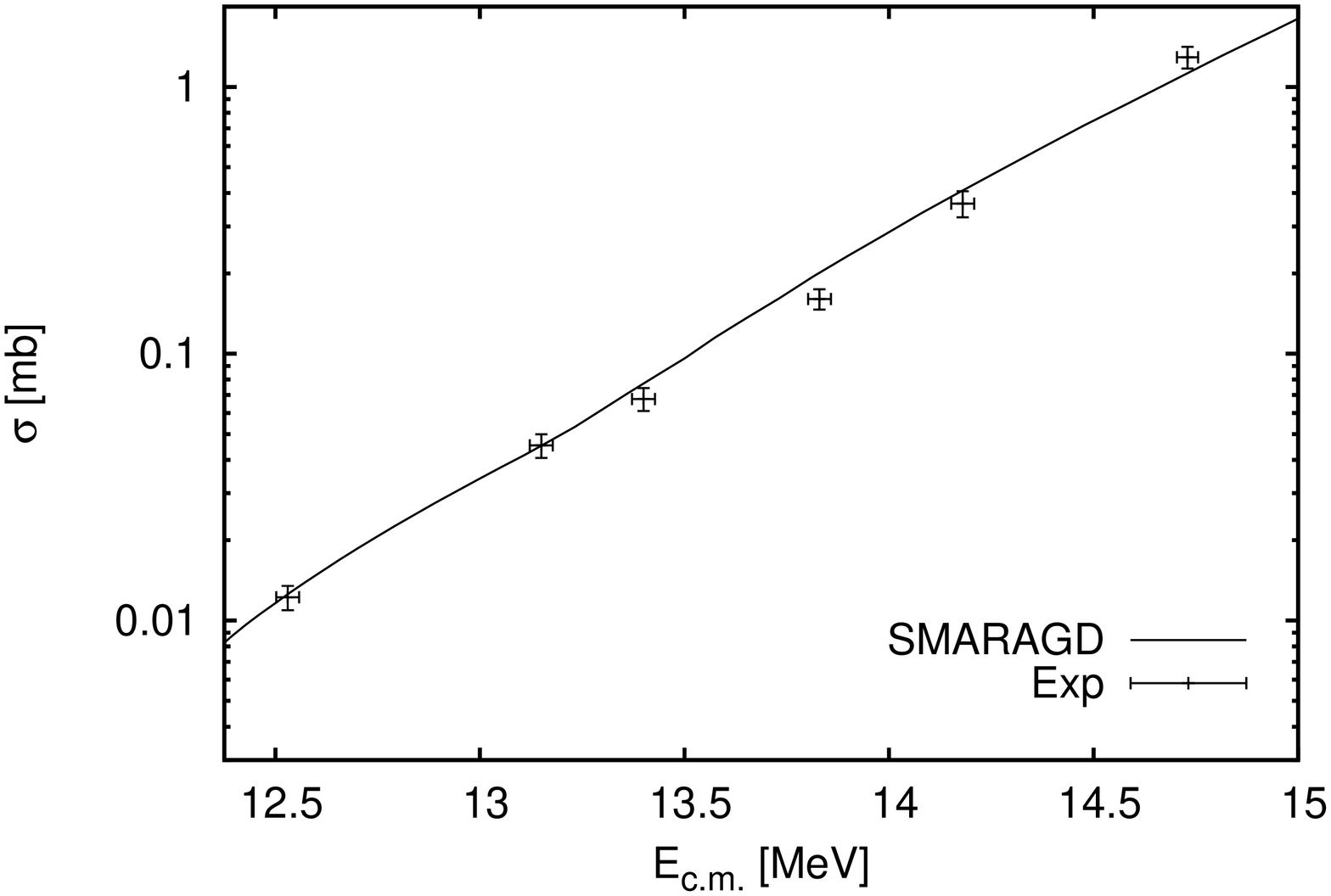}\\
\includegraphics[angle=-90,width=0.65\columnwidth]{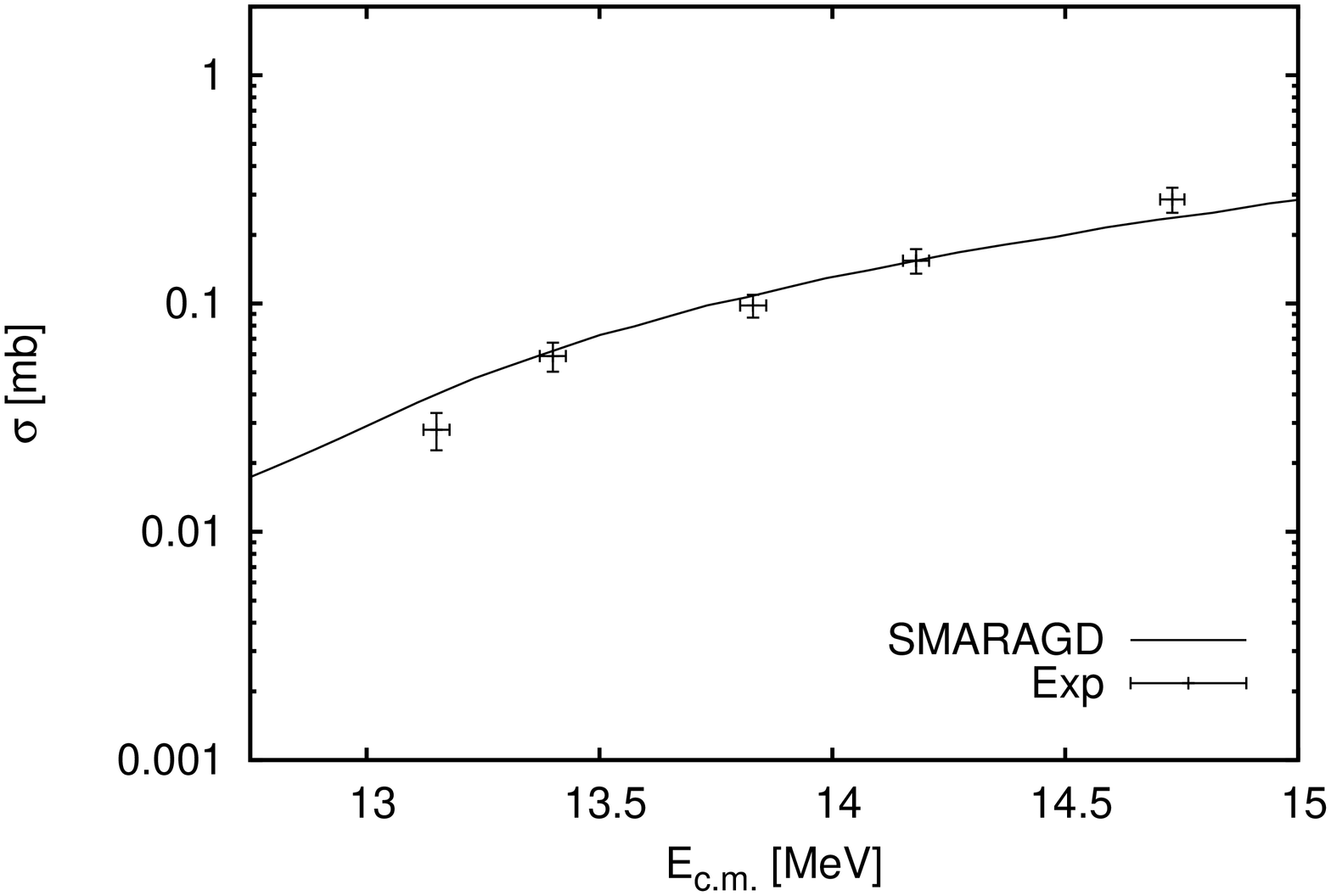}
\end{center}
\caption{Experimental cross sections \cite{annediss} of $^{168}$Yb($\alpha$,n)$^{171}$Hf (top) and $^{168}$Yb($\alpha$,$\gamma$)$^{171}$Hf (bottom) are compared to predictions with the code SMARAGD using the standard potential \cite{mcf}.}
\label{fig:yb168}
\end{figure}

\begin{figure}[h]
\begin{center}
\includegraphics[angle=-90,width=0.8\columnwidth]{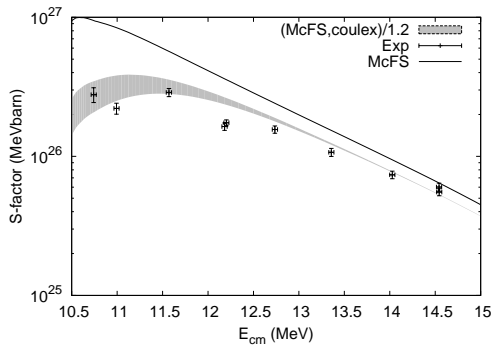}
\end{center}
\caption{Experimental S-factors for $^{141}$Pr($\alpha$,n)$^{144}$Pm \cite{pr141} (Exp) are compared to predictions using the standard potential \cite{mcf} (McFS) and correction for Coulomb excitation (McFS, Coulex), with additional 20\% renormalization. The uncertainty introduced by the $B$(E2) values is shown by the shaded region.}
\label{fig:pr141}
\end{figure}

\begin{figure}[h]
\begin{center}
\includegraphics[angle=-90,width=0.8\columnwidth]{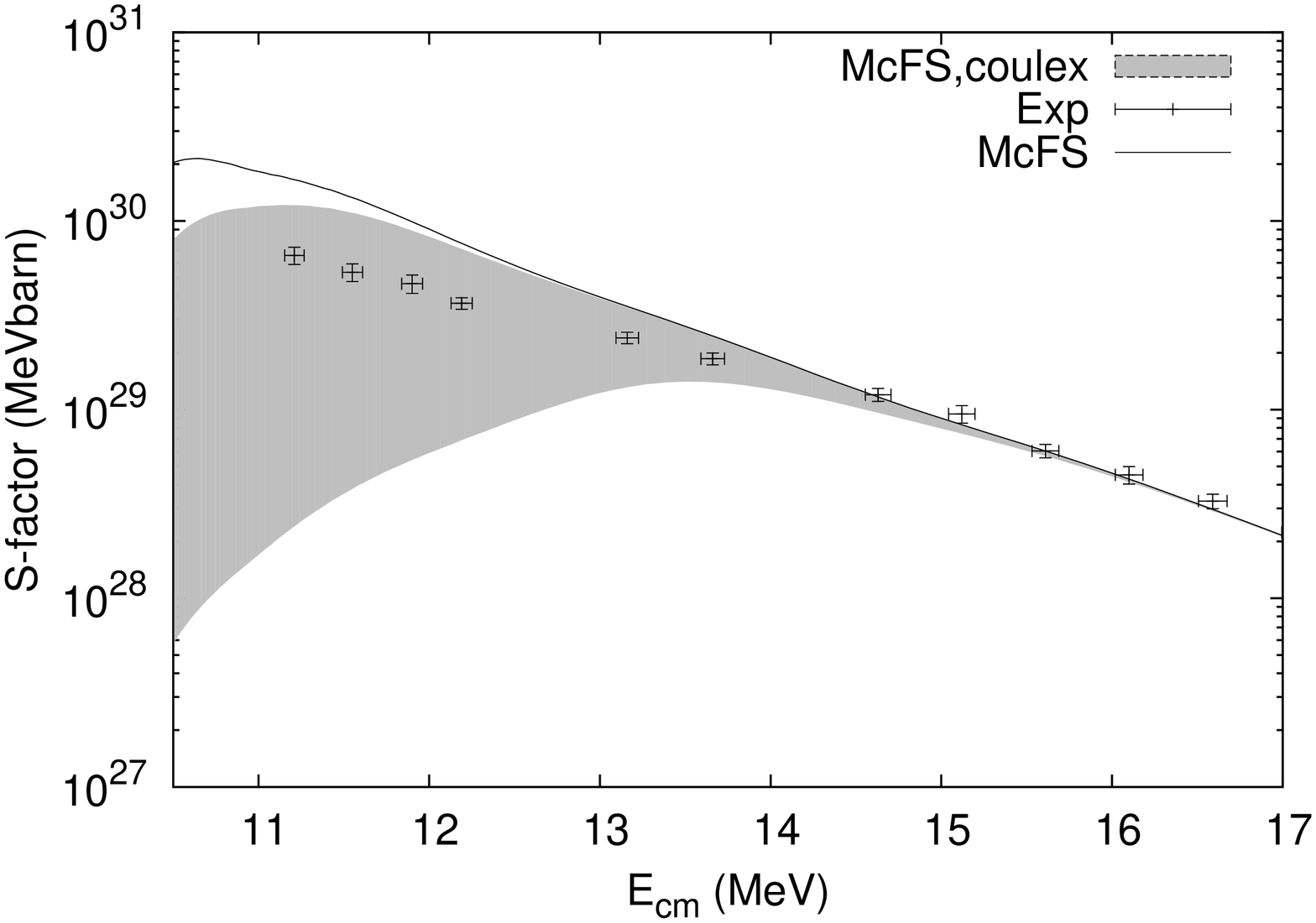}
\end{center}
\caption{Experimental S-factors for $^{169}$Tm($\alpha$,n)$^{172}$Lu \cite{tm169} (exp) are compared to predictions using the standard potential \cite{mcf} (McFS) and correction for Coulomb excitation (McFS, Coulex). The uncertainty introduced by the $B$(E2) values is shown by the shaded region.}
\label{fig:tm169}
\end{figure}

The approach outlined above should also remain valid when applied to other reactions. Due to the scarcity of suitable data, there are only few cases to be checked.
Already without inclusion of low-energy Coulex, very good agreement was found between predictions and data for $^{130,132}$Ba($\alpha$,n) \cite{ba130}. Despite the presence of low-lying 2$^+$ states, this remains so when including Coulex because its cross section $\sigma_\ell^\mathrm{Coulex}$ remains small compared to $\sigma_\ell^\mathrm{form}$ in the investigated energy range for the relevant partial waves. Another example for such a case is shown in Fig.\ \ref{fig:yb168},
where recent data \cite{annediss} for $^{168}$Yb($\alpha$,n)$^{171}$Hf and $^{168}$Yb($\alpha$,$\gamma$)$^{171}$Hf are compared to predictions. The excellent reproduction of the ($\alpha$,n) data -- which mainly depends on the correct description of the $\alpha$-widths \cite{sensipaper} -- shows that the standard potential \cite{mcf} fares well. The slight deviations from the data found in the ($\alpha$,$\gamma$) reaction must be due to the modeling of the $\gamma$- and/or neutron-widths but are not astrophysically relevant, as the $\alpha$-capture cross sections depend only on the $\alpha$-width at astrophysical energies \cite{sensipaper}.
Two further cases are shown in Figs.\ \ref{fig:pr141} and \ref{fig:tm169}, the reactions $^{141}$Pr($\alpha$,n)$^{144}$Pm and $^{169}$Tm($\alpha$,n)$^{172}$Lu, respectively. In both cases, the increasing deviation found for decreasing energy can be nicely explained by Coulex. A large uncertainty, however, remains in the $B$(E2) values which are experimentally not well determined for odd nuclei (or nuclei with g.s.\ other than 0$^+$). The prediction for $^{141}$Pr($\alpha$,n) may need a small modification of the optical potential, it is 20\% too high. But this is considerably lower than the usually assumed uncertainties in astrophysical rate predictions. The large uncertainty stemming from the $B$(E2) value does not allow to draw a final conclusion on $^{169}$Tm($\alpha$,n) but it seems that it may be feasible to reproduce the energy dependence of the data without change in the optical potential.

\begin{table*}
\caption{Stellar $^{144}$Sm($\alpha$,$\gamma$)$^{148}$Gd reactivities at a plasma temperature 2.5 GK from different sources, obtained with different codes and different types of optical $\alpha$+nucleus potentials. Also shown are the final $^{146}$Sm/$^{144}$Sm production ratios $\mathcal{R}$ obtained for different $^{144}$Sm($\alpha$,$\gamma$)$^{148}$Gd rates (and their reverse rates) in two models of the ccSN of a 25 $M_\odot$ star (ccSN-A \cite{ray95,sm144} and ccSN-B \cite{rhhw02,heg}) and a SNIa model \cite{travWDproc}. The values obtained with the optical potential of this work are given on the last line.}
\begin{tabular}{cccccc}
\hline
Type  &Code& \multicolumn{1}{c}{Reactivity} & \multicolumn{3}{c}{$\mathcal{R}$} \\
&& \multicolumn{1}{c}{(cm$^3$ s$^{-1}$ mole$^{-1}$)} & ccSN-A & ccSN-B & SNIa \\
\hline \hline
Equivalent Square Well \cite{tru72}   & CRSEC \cite{woohow78} & $3.8\times 10^{-15}$     &&&         \\
Folding (real), Woods-Saxon (imag.)                  & SMOKER\footnotemark[1] \cite{raulett} & $1.3\times 10^{-15}$     &&&       \\
Woods-Saxon \cite{mcf}                          & NON-SMOKER\footnotemark[1] \cite{nonsmoker} & $1.9\times 10^{-15}$   & 0.19 & 0.15 & 0.32   \\
Woods-Saxon \cite{mcf}                          & SMARAGD, this work & $2.4\times 10^{-14}$    & 0.11 & 0.06 &       \\
Energy-dep.\ Woods-Saxon \cite{sm144}             & MOST\footnotemark[1] \cite{sm144}, SMOKER\footnotemark[1] \cite{raulett} & $1.3\times 10^{-16}$  & 0.44 & 0.39 &      \\
Energy-dep.\ Woods-Saxon \cite{sm144} & SMARAGD, this work &$2.2\times 10^{-15}$   & 0.19 & 0.15 &         \\
Woods-Saxon \cite{mcf}, scaled $\alpha$-width    & SMARAGD, this work & $1.2\times 10^{-14}$   & 0.13 & 0.08 &           \\
\hline \hline
\end{tabular}
\label{table}
\footnotetext[1]{
The codes SMOKER, NON-SMOKER, MOST used the same routine to calculate Coulomb barrier penetration.
}
\end{table*}


To assess the impact on the stellar $^{148}$Gd($\gamma$,$\alpha$) rate it should be recalled that Coulex acts in the \textit{entrance} channel but the $\alpha$-emission channel should be unaffected. This is also the reason why an optical potential accounting for compound formation \textit{without} including Coulex in its absorptive part has to be used. Only such a potential can then be applied to $\alpha$-emission. (Detailed balance then applies to transitions obtained with such a potential.) This is not the potential that would be obtained by $\alpha$-scattering. If it were possible to perform an $\alpha$-scattering experiment at such low energy and extract an optical potential without correcting for Coulex, this potential would include both compound formation and Coulex in its absorptive part but no information on how to distribute the flux across the two possibilities.
The result \textit{without} Coulex also has to be used for computing the stellar reactivity $R_{\alpha \gamma}=N_\mathrm{A} \langle \sigma v \rangle ^*_{\alpha \gamma}$ for $^{144}$Sm($\alpha$,$\gamma$), which then can be converted to the ($\gamma$,$\alpha$) reactivity $R_{\gamma \alpha}$. Since the $\alpha$-width had to be reduced to reproduce the data after Coulex was applied (see Fig.\ \ref{fig:sm144}), it also has to be reduced in the original result without Coulex.


Table \ref{table} compares the stellar reactivities for $^{144}$Sm($\alpha$,$\gamma$) obtained with different codes and different potentials, as used in astrophysical applications. It should be noted that the codes also use different treatments of Coulomb barrier penetration and recently only the implementation in the SMARAGD code has been shown to be adequate for $\alpha$-transmission far below the Coulomb barrier. The final SMARAGD prediction (last line in Table \ref{table}) is higher than all previous estimates used in stellar models and in particular higher by two orders of magnitude than the value obtained by directly fitting the experimental results \cite{sm144}. This leads to a reduced isotope ratio $\mathcal{R}$. The actual value varies slightly between stellar models and also depends on the $^{148}$Gd($\gamma$,n) used, as this reaction competes with $^{148}$Gd($\gamma$,$\alpha$). Table \ref{table} also shows $\mathcal{R}$ obtained from postprocessing of three different models, using different $^{148}$Gd($\gamma$,$\alpha$) rates (and their reverses). Two models use trajectories from ccSN explosions of 25 solar masses ($M_\odot$) progenitor stars with solar metallicity: one similar to \cite{ray95} (ccSN-A) and a new model similar to \cite{rhhw02} but with initial solar abundances from \cite{lodd09} (ccSN-B). The value for the SNIa is taken from \cite{travWDproc}. All of the calculations use the predicted $^{147}$Gd(n,$\gamma$)$^{148}$Gd rate from \cite{nonsmoker}. Using an open-box model for galactic chemical evolution and neglecting any further free-decay and mixing timescales before inclusion into the early solar system (ESS), a range of $0.2\leq \mathcal{R} \leq 0.23$ is permitted by the $^{146}$Sm/$^{144}$Sm ratio in the ESS inferred from meteoritic data \cite{raupreview,kino}. Slightly higher values of $\mathcal{R}$ can be accommodated by making further assumptions on additional timescales during which the produced $^{146}$Sm decays before being incorporated into ESS solids. Although the ccSN isotope ratios $\mathcal{R}$ vary due to model differences, they are too low to fall into the permitted range. There is no calculation available for SNIa with the new potential but if the reduction in the ratio is of the same order as found for the ccSN models, then the new ratio could well be within the permitted range. 

The new, low value of $\mathcal{R}$ challenges explosive nucleosynthesis models as well as investigations in galactic chemical evolution and the formation of solids in the ESS. Further studies in both astrophysics and nuclear physics, however, are required to determine the actual value. Details in the stellar modelling and the used $^{12}$C($\alpha$,$\gamma$)$^{16}$O rate \cite{ray95} will impact the resulting ratio. Moreover, contributions from massive stars with different masses and initial composition are superposed during galactic evolution. Here, we only showed examples for 25 $M_\odot$ stars. Finally, the actual seed distribution which is photodisintegrated does not influence $\mathcal{R}$,
since both $^{146}$Sm and $^{144}$Sm originate from the photodisintegration of $^{148}$Gd. Not only the peak temperature reached in a zone, however, but also the temperature evolution, i.e., how much time is spent at a given temperature, impacts the final ratio. A higher temperature favors ($\gamma$,n) with respect to ($\gamma$,$\alpha$) and increases $^{146}$Sm production \cite{woohow}. The production ratio thus also depends on the expansion timescale, higher explosion temperatures are relevant with shorter timescales. The expansion is different in different ccSN models and it may be very different for SNIa. Following the expansion of the expanding hot fragments of both ccSN and SNIa -- and thus of their actual nucleosynthesis -- requires a detailed understanding of the explosion and accurate, high-resolution hydrodynamic modelling before final conclusions can be drawn.

Nuclear experiments can help to test the low-energy Coulex effect introduced here. For improved results, the $B$(E2) values for odd nuclei have to be determined with higher precision. In addition, if possible, a simultaneous detection of the $\gamma$-emission from the excited target nucleus state while performing a reaction experiment could directly indicate the action of Coulex. Complementary measurements of low-energy $\alpha$-absorption and -emission (difficult for $^{144}$Sm, obviously, but feasible for other test cases) should show a difference in the two directions, not accountable for by straightforward application of detailed balance. In this context it is interesting to note that (n,$\alpha$) experiments on $^{143}$Nd and $^{147}$Sm find an overprediction by a factor of 3 \cite{nalpha0,nalpha1,koe04,gle09}. This is fully consistent with the required renormalization found here for $^{144}$Sm+$\alpha$, after correction for Coulex. Finally, the isotope ratio $\mathcal{R}$ also depends on the $^{147}$Gd(n,$\gamma$)$^{148}$Gd rate which is unconstrained by experiment.

This work is partly supported by the Hungarian Academy of Sciences, the ESF EUROCORES programme EuroGENESIS, the ENSAR/THEXO collaboration within the 7th Framework Programme of the EU, the European Research Council, and the Swiss NSF.

\end{document}